\documentstyle[epsfig,12pt]{article}
\newcommand{\ur}[1]{(\ref{#1})}

\newcommand{\eq}[1]{eq.~(\ref{#1})}

\newcommand{\Eq}[1]{Eq.~(\ref{#1})}
\newcommand{\Eqs}[2]{Eqs.(\ref{#1}, \ref{#2})}

 \def\Sp{\mbox{Sp}}
 \def\Tr{\mbox{Tr}}
  \newcommand{\la}[1]{\label{#1}}
  \def\beq{\begin{equation}}
  \def\eeq{\end{equation}}

\begin{document}
\thispagestyle{empty}
\vskip 2.5true  cm
\begin{center}
{\Large\bf Potential Energy of Yang--Mills Vortices \\
\vskip .5true cm
in Three and Four Dimensions }
\\
\vskip 1.5true cm

{\large\bf Dmitri Diakonov}
\\[1cm]
{\it NORDITA, Blegdamsvej 17, 2100 Copenhagen \O, Denmark}\\
and\\
{\it Petersburg Nuclear Physics Institute, Gatchina,
St.Petersburg 188350, Russia} \\
\vskip .5true cm
E-mail: diakonov@nordita.dk
\end{center}
\vskip 1.5true cm
\begin{abstract}
\noindent We calculate the energy of a Yang--Mills vortex as function
of its magnetic flux or, else, of the Wilson loop surrounding the
vortex center. The calculation is performed in the 1-loop
approximation. A parallel with a potential as function of the Polyakov
line at nonzero temperatures is drawn. We find that quantized $Z(2)$
vortices are dynamically preferred though vortices with arbitrary
fluxes cannot be ruled out.
\end{abstract}

\vskip 2true cm

The hypothesis that $Z(N_c)$ vortices are responsible for
confinement has several attractive features. First, one immediately
gets the area law for Wilson loops in the fundamental representation
of the $SU(N_c)$ gauge group, even under the simplest assumption
that vortices are non-interacting and therefore the number of
vortices piercing a given loop is Poisson-distributed \cite{G1}.
Furthermore, Wilson loops in any representation transforming under the
group center $Z(N_c)$ will have the same string tension as in the
fundamental representation, while the `center-blind' representations
will have, asymptotically, zero string tension. This is what is expected
from screening by gluons. The temporal approximate Casimir scaling of
the string tensions may be probably explained by the finite sizes of
the vortex cores \cite{G2}.

Second, heating the $d=3+1$ Yang--Mills system towards the
high-temperature $d=3$ case and then heating $d=2+1$ system eventually to
the $d=2$ case, one does not observe abrupt changes in the {\em spatial}
string tensions. This continuity as one goes from four to two dimensions
favours $d=2$ objects as being basic for confinement in all the above
cases \footnote{In $d=2$ pure Yang--Mills theory the confinement is
trivial in axial gauges and may seem to have no relation to vortices.
However, it is known that at least on compact $d=2$ manifolds
the theory is exactly equivalent to a sum over vortices, see a recent
paper \cite{d2vort} and references therein.}. In three dimensions
vortices form closed lines; in four dimensions they form closed
surfaces.

For vortices to be physical objects and not artifacts of a
regularization, their cores should be finite in physical units, i.e. to
be of the order of $1/\Lambda\simeq M^{-1}\exp(3\cdot 8\pi^2/11 N_c g^2)$
in $d=4$ and of the order of $1/g^2_3$ in $d=3$. Correspondingly, the
energy of a vortex per unit length should be of the order of $g_3^2$ in
$d=3$, and the energy per unit surface should be of the order of
$\Lambda^2$ in $d=4$.  First indications that in might indeed be the case
came recently from lattice studies using the smoothing procedure
\cite{G3}.

In order to reveal `thick' vortices theoretically, one has first of all
to integrate out the high-momentum components of the Yang--Mills field.
The resulting effective action may then have a stable saddle point
of a vortex type, with the core size remaining finite as one takes
the UV cutoff to infinity. When and if it happens, one can speak
of vortices as physical entities, and further on to try to build a
theory of those extended objects in the vacuum.

The main dynamical variable of a vortex is the azimuthal component
of the Yang--Mills field $A_\phi^a(\rho)=\epsilon_{\alpha\beta}
n_\alpha A_\beta^a$, where $n_\alpha$ is a unit vector in the plane
transverse to the vortex. One can always choose a gauge where $A_\phi$
is independent of the azimuth angle $\phi$. Generally speaking, it
implies that the radial component $A_\rho(\rho)\neq 0$, however, we
shall neglect this component as it can be always reconstructed from
gauge invariance by replacing $\partial_\rho\to
\partial_\rho\delta^{ab} + f^{acb}A_\rho^c$. A circle Wilson loop lying
in the transverse plane and surrounding the vortex center is then

\beq
W_{\frac{1}{2}}(\rho)=\frac{1}{2}
\Tr\;{\rm P}\exp i \oint\rho A_\phi^a t^a d\phi=
\cos[\pi\mu(\rho)],\;\;\;\;\;
\mu(\rho)=\rho\sqrt{A_\phi^a(\rho) A_\phi^a(\rho)},
\la{W1/2}\eeq
taking for simplicity the fundamental representation of the $SU(2)$
gauge group. For an arbitrary representation of $SU(2)$, labelled by
spin $J$, one has

\beq
W_J(\rho)=\frac{1}{2J+1}\frac{\sin[(2J+1)\pi\mu(\rho)]}
{\sin[\pi\mu(\rho)]}.
\la{WJ}\eeq
If $\mu(\rho)\to integer$ at large distances $\rho$ from the vortex
core, the Wilson loop $W_J(\rho)\to (-1)^{2J}$. This is the definition
of the $Z(2)$ vortex. We shall see that integer values of $\mu(\infty)$
are dynamically preferred.

Neglecting all components of the Yang--Mills field except the
essential azimuthal one, the classical action of the vortex becomes

\beq
\int\!d^dx\frac{(F^a_{\mu\nu})^2}{4g_d^2}
=\int\!d^{d-2}x_\parallel\!\int\!\!d^2x_\perp\frac{(B^a_\parallel)^2}
{2g_d^2}=\frac{1}{2g_d^2}\int\!d^{d-2}x_\parallel\;
2\pi\!\int_0^\infty\!\!d\rho\:\rho
\left[\frac{1}{\rho}\partial_\rho(A_\phi^a\rho)\right]^2.
\la{classaction}\eeq

To get the effective action for $\mu(\rho)$ we integrate over quantum
fluctuations about a given background field $A_\phi^a(\rho)\rho$
considered to be a slowly varying field with momenta up to certain
$k_{min}$. Accordingly, quantum fluctuations have momenta from
$k_{min}$ up to the UV cutoff $k_{max}$. Writing $A_\mu=\bar A_\mu +
a_\mu$ where $\bar A_\mu=\delta_{\mu\phi}A_\phi^a(\rho)$ we expand
$F^2_{\mu\nu}(A+a)$ in the fluctuation field $a_\mu$ up to the second
order appropriate for the 1-loop calculation. The term linear in $a_\mu$
vanishes due to the orthogonality of high and low momenta. The quadratic
form for $a_\mu$ is the standard

\beq
W^{ab}_{\mu\nu}=-[D^2(\bar A)]^{ab}\delta_{\mu\nu}-2f^{acb}
F^c_{\mu\nu}(\bar A),
\la{quadrform}\eeq
if one imposes the background Lorentz gauge condition,
\beq
D_\mu^{ab}(\bar A) a_\mu^b = 0,\;\;\;\;\;
D_\mu^{ab}(\bar A) = \partial_\mu\delta^{ab}+f^{acb}\bar A_\mu^c.
\la{backgauge}\eeq
The effective action is
\beq
S_{eff}[\bar A] = \frac{1}{2}\ln\det(W_{\mu\nu})
-\ln\det(-D_\mu^2).
\la{effact}\eeq

In the presence of dynamical fermions in the fundamental representation
the Dirac determinant should be added:

\beq
-\ln\det(\nabla_\mu\gamma_\mu)=-\frac{1}{2}\ln\det\left(
\nabla_\mu^2-\frac{i}{2}[\gamma_\mu\gamma_\nu]F_{\mu\nu}^at^a\right),
\;\;\;\;\;\;\nabla_\mu=\partial_\mu-i\bar A_mu^at^a.
\la{dirac}\eeq

The effective action may be expanded in powers of the (covariant)
derivatives of the background field $\bar A_\mu$. We are now
interested in the first nontrivial term of this expansion, namely in
the zero-derivative term, that is in the effective {\em potential} as
function of the flux $\rho A_\phi$. The next term with two derivatives
will renormalize \eq{classaction}, and there will be, generally speaking,
further terms.  In the zero-derivative order $F_{\mu\nu}(\bar A)=0$,
therefore $\det(W_{\mu\nu})=d\cdot\det(-D^2)$, where $d$ is the full
space dimension. With $A_\phi$ being the only nonzero component of the
background field, it is natural to write down $D^2$ in the cylindric
coordinates:

\beq
D^2=
\frac{1}{\rho}\frac{\partial}{\partial\rho}\rho
\frac{\partial}{\partial\rho}+\frac{1}{\rho^2}
\left(\frac{\partial}{\partial\phi}+f^{acb}A^c_\phi\right)^2
+\partial^2_3+...+\partial^2_d.
\la{cyl}\eeq
The effective action \ur{effact} for slowly varying $\rho A_\phi$
is

\beq
S_{eff}[A_\phi]
=-\left(\frac{d}{2}-1\right)
\int_{s_{min}}^{s_{max}}\!\frac{ds}{s}\;\Sp\;\exp(sD^2)
=\int\!\!d^{d-2}x_\parallel\!\int\!\!d^2x_\perp\;V^{(d)}(\mu),
\;\;\;\;\;\;\mu=\rho\sqrt{A_\phi^a A_\phi^a},
\la{bosact}\eeq
where we have introduced the `vortex potential energy' in $d$
dimensions $V^{(d)}(\mu)$.
Here $\Sp$ denotes the functional and the colour traces, and integration
over the proper time $s$ has been introduced. The limits $s_{min}\simeq
1/k^2_{max}$ and $s_{max}\simeq 1/k^2_{min}$ are gauge-invariant UV and
IR cutoffs, respectively. In saturating the functional trace one can
use any complete set of functions. Naturally, one uses the plane wave
basis for the longitudinal components $x_3...x_d$, and the polar basis
$\exp(im\phi)f_{k,m}(\rho)$ for the transverse ones, where the functions
$f_{k,m}(\rho)$ must form a complete set, see below. In what follows we
shall consider only the $SU(2)$ case. The transverse part of the
covariant Laplacian \ur{cyl}, after acting on the polar harmonics
$\exp(im\phi),\;\;m=\; integer$, has three eigenvalues as a $3\times
3$ colour matrix:

\beq
\frac{1}{\rho}\frac{\partial}{\partial\rho}\rho
\frac{\partial}{\partial\rho}-\frac{1}{\rho^2}\cdot
\left\{\begin{array}{c}m^2,\\(m+\mu)^2,\\(m-\mu)^2.\end{array}
\right.
\la{eigen}\eeq
The first one cancels out when one subtracts the free determinant
(without the external field); the last two are actually coinciding,
given that $m$ ranges from minus to plus infinity. The differential
operator (\eq{eigen}, last line) has eigenfunctions
$J_{\pm(m-\mu)}(k\rho)$ with eigenvalues $-k^2$. The index of Bessel
functions must be non-negative to ensure regularity at the origin. We
thus choose the functions
$F_{m,k}(\phi,\rho)=\exp(im\phi)J_{|m-\mu|}(k\rho)$ as a complete
functional basis in the transverse plane:

\beq
\frac{1}{2\pi}\sum_{m=-\infty}^{+\infty}\int_0^\infty dk\:k\:
F_{m,k}(\phi,\rho)F_{m,k}^\star(\phi^\prime,\rho^\prime)
=\frac{1}{\rho}\delta(\rho-\rho^\prime)
\left.\delta(\phi-\phi^\prime)\right|_{{\rm mod}\;2\pi}.
\la{complete}\eeq
These functions satisfy also the ortho-normalization condition:

\beq
\frac{1}{2\pi}\int_{0}^{2\pi}d\phi\int_0^\infty d\rho\:\rho\:
F_{m,k}(\phi,\rho)F_{m^\prime,k^\prime}^\star(\phi,\rho)
=\frac{1}{k}\delta(k-k^\prime)\;\delta_{mm^\prime}.
\la{orthon}\eeq
\Eqs{complete}{orthon} can be checked using the formula
6.541.1 from Gradshteyn and Rhyzhik \cite{GR}. \Eq{bosact} can be thus
rewritten as

\[
V^{(d)}(\mu)=-(d-2)\int\frac{dp_3...dp_d}{(2\pi)^{d-2}}
\int_{s_{min}}^{s_{max}}\frac{ds}{s}\int_0^\infty k\:dk\:
\exp[-s(k^2+p_3^2+...+p_d^2)]\]
\beq
\cdot\frac{1}{2\pi}\sum_{m=-\infty}^\infty
\left[J^2_{|m-\mu|}(k\rho)-J^2_{|m|}(k\rho)\right].
\la{bosact1}\eeq
We see that the dependence on the flux $\mu$ enters only through
the indices of the Bessel functions and that the potential is
explicitly periodic in $\mu$ with period 1. Integration over
momenta can be easily performed using eq. 6.633.2 from \cite{GR},
yielding

\beq
V^{(d)}(\mu)=-(d-2)\int_{s_{min}}^{s_{max}}\frac{ds}{s}
\left(\frac{1}{4\pi
s}\right)^{\frac{d}{2}}\exp\left(-\frac{\rho^2}{2s}\right)
\sum_{m=-\infty}^\infty
\left[I_{|m-\mu|}\left(\frac{\rho^2}{2s}\right)
-I_{|m|}\left(\frac{\rho^2}{2s}\right)\right].
\la{bosact2}\eeq
We next sum over $m$ using the integral representation for modified
Bessel functions, eq. 8.431.5 of \cite{GR}. To write explicit formulae
we imply that $\mu\in(0,\:1)$. If $\mu$ is outside this interval
$V^{(d)}(\mu)$ should be continued by periodicity. We get

\beq
V^{(d)}(\mu)=\frac{d-2}{\left(2\pi\rho^2\right)^{\frac{d}{2}}}
\cdot\frac{\sin(\pi \mu)}{\pi}\!\int_0^\infty\!dx
\frac{\cosh\left(\frac{1}{2}-\mu\right)x}{\cosh\frac{x}{2}}
\int_{t_{min}}^{t_{max}}\!dt\:t^{\frac{d}{2}-1}
\exp\left[-t(\cosh x+1)\right],
\la{bosact3}\eeq
where we have introduced a new variable $t=\rho^2/(2s)$ instead of $s$.
Correspondingly, the integration limits become $t_{min}=\rho^2/2s_{max}$
and $t_{max}=\rho^2/2s_{min}$. It should be stressed that the above
expression is finite in the limit when one removes both the UV cutoff
($t_{max}\to\infty$) and the IR cutoff ($t_{min}\to 0$). In this limit
both integrations in \ur{bosact3} become elementary, and we finally get:

\[
V^{(d)}(\mu)=\frac{d-2}{\left(\pi\rho^2\right)^{\frac{d}{2}}}\cdot
\frac{\sin(\pi \mu)}{\pi}\cdot\frac{\Gamma\left(\frac{d}{2}\right)
\Gamma\left(\frac{d}{2}+\mu\right)\Gamma\left(\frac{d}{2}+1-\mu\right)}
{\Gamma(d+1)}\]
\beq
=\;\;\left\{\begin{array}{cc}
\left.\frac{1}{\rho^4}\frac{1}{12\pi^2}\;
\mu(1-\mu^2)(2-\mu)\right|_{{\rm mod}\;1}\;\;\;\;
{\rm for}\;\;d=4,\\
\\
\left.\frac{1}{\rho^3}\frac{1}{96}\;\frac{\tan(\pi\mu)}{\pi}
(1-4\mu^2)(3-2\mu)\right|_{{\rm mod}\;1}\;\;\;\;
{\rm for}\;\;d=3.
\end{array}\right.
\la{bosact4}\eeq
At $d=2$ the vortex potential is identically zero as there
are no transverse gluons in two dimensions.
Though the potential is given just by a polynomial (in case $d=4$),
it is actually a periodic function of $\mu$ with a unit period, as
seen explicitly from \eq{bosact1}. At integer $\mu$ the potential
is zero but has a jump in the first derivative; it is depicted in
Fig.1.

\begin{figure}
\centerline{\epsfxsize10.0cm\epsffile{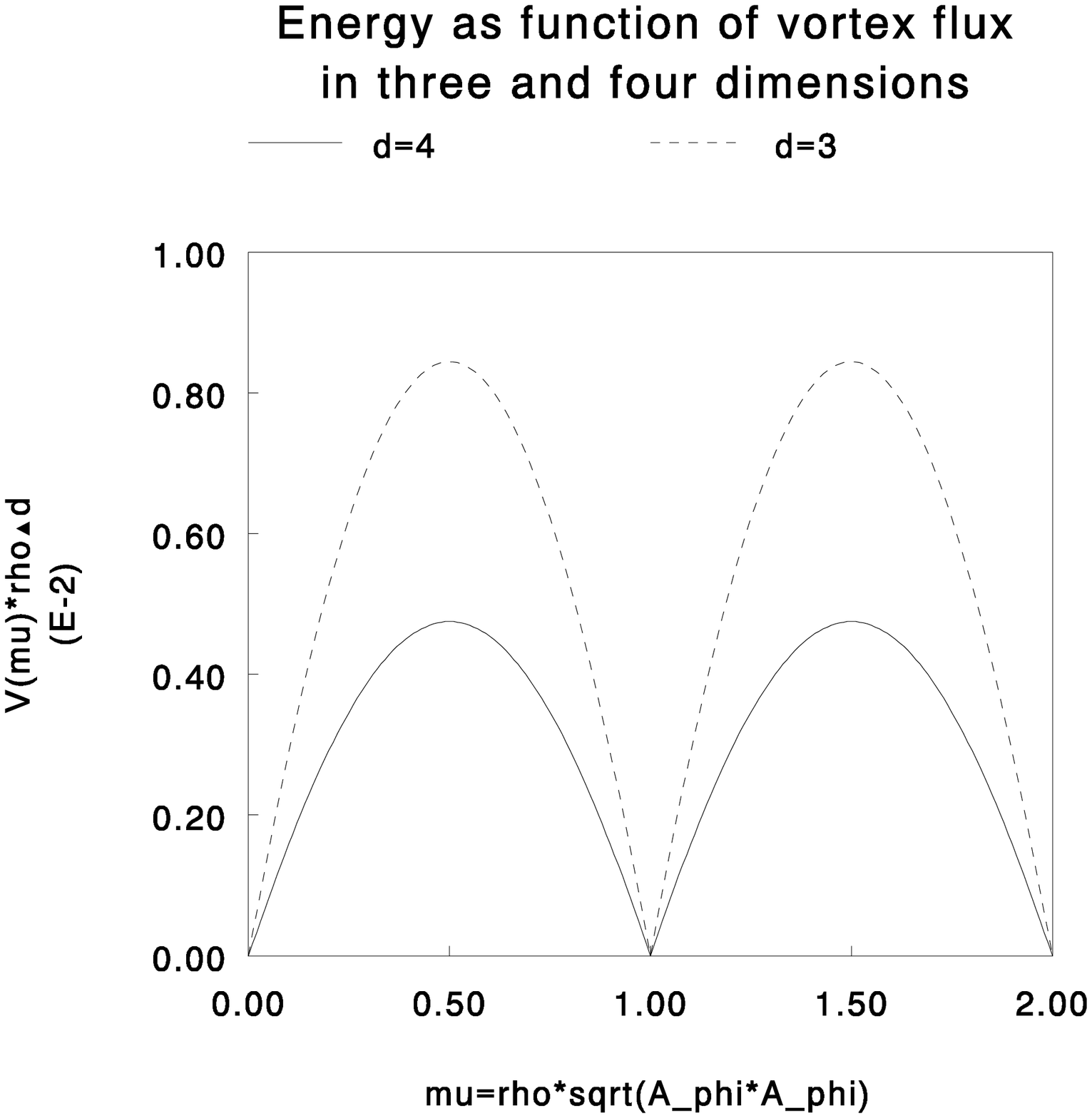}}
\caption[]{The vortex potential, \protect\eq{bosact4}.}
\end{figure}

The problem we have solved has certain similarity with that of the
potential energy as function of the $d$-th component of the Yang--Mills
field $A^a_{\mu=d}$, in case of nonzero {\em temperatures}. In that
problem one integrates over all Matsubara frequences in the background
of a static $A^a_d$, where $d=4$ for $3+1$ dimensions, and $d=3$ for
$2+1$ dimensions. The two problems are ideologically similar, only
the topology is different: in the nonzero-temperature case the topology
is that of a cylinder $S^{(1)}\times R^{(d-1)}$, with the compact
dimension in the `temperature' direction, while in the problem
considered here the topology is that of a plane with a deleted point at
the origin, $R^{(2)}\setminus\{0\}\times R^{(d-2)}$. The role
of Matsubara frequencies is played by the polar harmonics
$\exp(im\phi)$. In the first case one finds the potential energy as
function of the Polyakov line winding in the compact dimension;
in the second case one finds the potential energy as function
of the Wilson loop winding around the point at the origin.
In both cases the potential energy is a periodic
function, however the calculations are easier in the
nonzero-temperature case as one can evaluate the functional trace
\ur{bosact} in the plane-wave basis. A simple calculation of
\eq{bosact} gives the following form of the nonzero-temperature
potential energy as function of $\nu=\sqrt{A_d^aA_d^a}/(2\pi T)$, valid
for any number of dimensions:

\beq
V^{(d)}(\nu)=(d-2)T\int\! \frac{d^{d-1}p}{(2\pi)^{d-1}}\;\ln
\frac{\cosh\frac{p}{T}-\cos 2\pi\nu}{\cosh\frac{p}{T}-1}.
\la{nonztemp1}\eeq
It is explicitly periodic in $\nu$. At $d=4$ the integration can be
performed with the help of eq. 3.533.3 \cite{GR}, and we get the
well-known result \cite{Weiss}:

\beq
V^{(4)}(\nu)=\left.\frac{(2\pi T)^4}{12\pi^2}\nu^2(1-\nu)^2
\right|_{{\rm mod}\;1}=\frac{T^2}{3}(A_4^a)^2+O(A^3).
\la{nonztemp4}\eeq
At $d=3$ the integral \ur{nonztemp1} cannot be expressed through
elementary functions; however it can be compactly written as

\[
V^{(3)}(\nu)=\frac{2T^3}{\pi}\sin^2\pi\nu\int_0^\infty\!dx
\frac{x^2
\cosh x}{\sinh x(\sinh^2 x+\sin^2\pi\nu)}
\]
\beq
=\pi T^3\nu^2\ln\frac{1}{\nu^2}+O(\nu^4)
=\frac{T}{4\pi}(A_3^a)^2\ln\frac{T^2}{(A_3^a)^2}+O(A^4).
\la{nonztemp3}\eeq
To our best knowledge, this is a new result.

It is worth mentioning that for $d=3$ the Debye mass is
infrared-divergent, hence the non-analiticity in the $\nu^2$ term. In
the case of the vortex potential \ur{bosact4} the infrared divergency
is even stronger: as a result the expansion of the potential starts
from the non-analytic $|\mu|$ term both for $d=3$ and 4. It is
interesting that in all cases above the non-analytic terms are due
to the contribution of zero Matsubara frequencies (in case of the
vortex ``zero frequency'' means the azimuthal harmonic with $m=0$).
Indeed, the contributions of zero frequencies to the `temperature'
potentials \ur{nonztemp4} and \ur{nonztemp3} are

\beq
V^{(4)}_{\omega=0}(\nu)=-\frac{(2\pi T)^4}{12 \pi^2}2\nu^3
\;+\;{\rm UV\;divergence},\;\;\;\;\;\;
V^{(3)}_{\omega=0}(\nu)=\pi T^3\nu^2\ln\frac{1}{\nu^2}
\;+\;{\rm UV\;divergence}.
\la{zerf1}\eeq
The $m=0$ contribution to the vortex potential \ur{bosact4} is
\beq
V^{(4)}_{m=0}(\mu)=\frac{1}{\rho^4}\frac{1}{12 \pi^2}(\mu-\mu^3)
\;+\;{\rm UV\;divergence}.
\la{zerf2}\eeq
These are exactly the non-analytic terms in all the cases. The UV
divergences are cancelled by contributions from nonzero frequencies.

If there are dynamical fermions in the problem, they should be treated
differently in the nonzero-temperature case and in the vortex one.
In the first case fermions are antiperiodic in the `temperature'
direction, hence one has to sum over half-integer Matsubara frequencies.
In the latter case fermion wave functions are periodic functions of the
azimuthal angle $\phi$. The resulting fermion contribution to the
vortex potential (in the fundamental representation) is obtained
from \eq{bosact4} by substituting $\mu\to\mu/2$ and multiplying the
result by $-2$. The fermion contribution is periodic in $\mu$
with period 2, and not 1 as in the boson case. It can be easily checked
that in the supersymmetric case, with Majorana gluinos belonging to the
adjoint representation, the fermion contribution cancels exactly with
the boson one, so that the vortex potential is zero.

Returning to the bosonic contribution to the effective vortex potential
\ur{bosact4} in the pure glue case, one concludes that integer values of
$\mu(\rho)$ are a must at $\rho\to 0$, otherwise the integral over
$\rho$,

\beq
{\rm potential\;energy} = 2\pi \int_0^\infty d\rho\:\rho\:
V^{(d)}(\mu),
\la{integral}\eeq
diverges. Integer values of $\mu(\rho)$ at $\rho\to\infty$ are,
clearly, energetically favourable though the energy \ur{integral}
remains finite at noninteger values of $\mu(\infty)$ as well because of
the convergence of the integration over $\rho$ at large $\rho$.  It
means that the quantized $Z(2)$ vortices are dynamically preferred but
noninteger fluxes are not altogether ruled out by the energetics.  This
should be contrasted to the case of nonzero temperatures where
quantized values of $A_d$ at spatial infinity are necessary to make the
energy finite. \\

I am grateful to Victor Petrov, Maxim Chernodub and Konstantin Zarembo
for useful discussions.


\begin{thebibliography}{99}

\bibitem{G1}
L.Del Debbio, M.Faber, J.Greensite and S.Olejnik, Phys. Rev. D55
(1997) 2298, hep-lat/961005

\bibitem{G2}
M.Faber, J.Greensite and S.Olejnik, Phys. Rev. D57 (1998) 2603,
hep-lat/9710039

\bibitem{G3}
L.Del Debbio, M.Faber, J.Giedt, J.Greensite and S.Olejnik,
Phys. Rev. D58 (1998) 094501, hep-lat/9708023; \\
R.Bertle, M.Faber, J.Greensite and S.Olejnik, hep-lat/9903023

\bibitem{d2vort}
L.Griguolo, Nucl. Phys. B547 (1999) 375, hep-th/9811050

\bibitem{GR}
I.S.Gradshteyn and I.M.Ryzhik, {\em Tables of Integrals, Sums, and
Products}, Academic Press (1980)

\bibitem{Weiss}
N.Weiss, Phys. Rev. D24 (1981) 475

\end{thebibliography}
\end{document}